\documentstyle[12pt,preprint,aps]{revtex}

\begin{document}

\newcommand{\cN}{{\cal N}}
\newcommand{\ov}{\overline{}}
\newcommand{\pa}{\partial }
\newcommand{\D}{\Delta }
\newcommand{\e}{\epsilon }

\preprint{ \vbox{\noindent COLO-HEP-469 \\
\vspace{-.15in} hep-th/0106174} }

\title{ Moduli stabilization and supersymmetry breaking\\ in effective
theories of strings}

\author{R. Brustein\protect\( ^{(1)}\protect \) and
S. P. de Alwis\protect\( ^{(2)}\protect \)}

\address{(1) Department of Physics, Ben-Gurion University, Beer-Sheva 84105, Israel
\\
 (2) Department of Physics, Box 390, University of Colorado, Boulder, CO 80309.\\
 \texttt{e-mail: (1) ramyb@bgumail.bgu.ac.il },
\texttt{(2)dealwis@pizero.colorado.edu}}

\maketitle

\begin{abstract}

The region of moduli space of string theories which is most
likely to describe the ``real world" is where the string coupling
is about unity and the  volume of extra compact dimensions is
about the same size as the string volume.  Here we map the
landscape of this ``central" region in a model-independent way,
assuming only that the string coupling and compact volume moduli
are chiral superfields of $\cN\!=\!1$ supergravity (SUGRA) in 4
dimensions, and requiring only widely accepted conditions:~that
the supersymmetry (SUSY) breaking scale is about the weak scale
and that the cosmological constant be of acceptably small
magnitude. We find that the superpotential has (in the
supersymmetric limit) a fourth order zero in the SUSY breaking
direction. The potential near the minimum is very steep in the
SUSY preserving directions, and very flat in the SUSY breaking
direction, consequently the SUSY breaking field has a weak scale
mass, while other moduli are heavy. We also argue that there will
be additional near by minima with a large negative cosmological
constant.

\end{abstract}


The most popular solution to the gauge hierarchy problem, i.e. the
stability of the tiny ratio \( m_{W}/M_{P}\sim 10^{-16} \)
between the weak scale and the Planck scale, envisages a \( \cN=1
\)  SUGRA  broken by some non-perturbative field theoretic
effects in some ``hidden" sector \cite{nilles}.  In this context
it is commonly held that one should compactify string theory on
Ricci flat manifolds such as a Calabi-Yau (CY)
spaces. Alternatively, in some brane world
scenarios  the hidden sector lives on one of the branes
and SUSY breaking is transmitted by gravitational effects to the
``visible" brane, on which our universe lives. However, the
potentials that are generated for the dilaton and the
compactification moduli are typically of the runaway form
\cite{dine} so that the theory prefers to go to the zero coupling
and decompactification limit. Similar problems are encountered in
theories in which SUSY is broken completely by string theoretic
effects, such as the Scherk-Schwarz mechanism where the
compactification explicitly breaks SUSY.

If one tries to stabilize moduli by introducing a more
complicated hidden sector as in race-track models \cite{kras},
then SUSY is broken at some intermediate scale (about
$10^{13}GeV)$. Alternatively, if one uses  string theoretic
mechanisms, such as the Scherk-Schwarz mechanism with the radius
of compactification stabilized by quantum effects, or models with
D branes and anti D branes at some orbifold fixed points or
D-branes at angles, SUSY is broken at the string scale. In such
mechanisms one invariably encounters what we have called the
``practical cosmological constant problem" (PCCP) \cite{bda}.
That is the problem of ensuring that to a given accuracy within a
given model the cosmological constant vanishes. This is
equivalent to the requirement that models should at the very
least allow for the possibility of a large universe to exist with
reasonable probability. This is not the same as requiring a
solution to the ``cosmological constant problem" \cite{weinberg}:
why is the cosmological constant so small in natural units?
\cite{foot0}

Thus, currently available stringy SUSY breaking and moduli
stabilization mechanisms are simply not viable. They do not even
allow a large universe, let alone finding an explanation for one.
One could take the point of view that the difficulties are
``technical", and that they will eventually be resolved when
computational technology improves, and therefore simply ignore
them. We believe that the difficulties are not technical, but
rather require some changes in the basic framework. Here we show
what modifications are needed in the input from string theory and
how they can  improve the situation significantly. We will not
have anything to say about the precise string mechanisms that may
lead to these modifications. However, the  point is that if
string theory is to solve even the PCCP, such a mechanism would
have to be found.

A generic resolution of the  PCCP  requires a continuously
adjustable constant in the potential. Consequently, in a $\cN=1$
SUGRA such a constant in the superpotential is required. So what
could be its origin? There is no mechanism known to us for its
generation from field theoretic effects, so if it exists, it must
originate directly from string theory. General solutions to the
string equations of motion have integration constants, but so far
only restricted choices of these constants have been made. In the
standard string phenomenology based on CY, or \( D\bar{D} \)
systems there is no room for a continuously adjustable constant
in the superpotential. It is the  Ricci flatness condition on CY
(or orbifold) compactfications that amounts to a restriction that
overly constrains the 4D moduli potential. On the other hand
brane world scenarios with compactification on (squashed)
spheres, such as type IIB on a squashed 5-sphere with \( D_{3} \)
branes \cite{Bremer} allow a less restrictive framework, with room
for an adjustable constant. In particular, it is possible to
choose integration constants to get flat brane solutions after
SUSY breaking\cite{dfi}. However it is still unclear whether such
models, which so far have only been constructed within IIB
supergravity, have a string theoretic ultra-violet completion.

Dualities, and the proposed unity of all string theories and 11D
SUGRA offer some new perspectives on the problem of moduli
stabilization and SUSY breaking. One of the well known lessons of
string dualities is that the strong coupling limit of one type of
perturbative string theory is another type of perturbative theory
(S-duality). Similarly a compactification on a `small' (compared
to the string scale) manifold is dual to  a different (or
sometimes the same) perturbative theory compactified on a
``large" manifold. Now, even in the absence of SUSY breaking one
would expect string theoretic non-perturbative (SNP) effects
coming from the various brane instantons\cite{bda,ovrut,moore}.
To the extent that their effects can be estimated (and this can
only be done in the various weak coupling limits) they give
runaway potentials which tend to take the theory to the extreme
weak or strong coupling limit and or the extreme
decompactification limit. For example, in the case of the
S-duality between the heterotic \( E_{8}\times E_{8} \) (HE)
theory and its strong coupling version the Horava-Witten (HW)
theory, the non-perturbative effects due to the F string and the
NS fivebrane instantons tend to give a runaway dilaton potential
which makes the dilaton roll down to weak coupling whereas in the
HW theory the corresponding dual effect originates from  the $M$
twobrane and the M fivebrane instantons which tend to push the
theory towards infinite radius for the eleventh dimension i.e. to
strong coupling.  A realistic phenomenolgy would thus require
that these effects are merely manifestations of the fact that the
investigations are done in the outer regions of moduli space and
that in the central region there exists an actual minimum of the
potentials for the various moduli\cite{bda}. This is the basic
idea in the concept of ``string universality''. Furthermore, we
have pointed out, in this context, that the stabilization of the
moduli was most probably a string non-perturbative (SNP) effect
while SUSY breaking could well be a field theoretic phenomenon.

The scenario that we propose for moduli stabilization and SUSY
breaking is then the following. The central region of moduli
space is parameterized by chiral superfields of D=4, $\cN=1$
SUGRA. They are all stabilized at the string scale by SNP effects
which allow a continuously adjustable constant in the
superpotential. SUSY is broken at an intermediate scale by field
theoretic effects that shift the stabilized moduli only by a
small amount from their unbroken minima. The cosmological
constant can be made to vanish after SUSY breaking by the
adjustable constant. We find that models of such a scenario
provide a surprisingly rich central region landscape. Although
our motivation for proposing this scenario originates from string
universality, our analysis and conclusions are valid whenever a
string theory can be approximated by an effective $\cN=1$ SUGRA
in the central region of moduli space.


We model the central region using two moduli, one parameterizes
the SUSY breaking direction, and the other is a representative of
the orthogonal directions. Keeping with tradition we name them
$S$ and $T$, but they are not necessarily the dilaton and volume
moduli of any of the perturbative string theories, rather, we
expect them to be functions of them and the other moduli (such as
shape moduli). We first study stabilization of moduli with
unbroken SUSY, as our scenario calls for, but instead of
considering the strict limit of unbroken SUSY, we introduce the
SUSY breaking scale \( \varepsilon \equiv m_{3/2}\sim 10^{-16} \)
(We use units in which the Planck mass \( M_{P}=1 \)), and
consider the dependence of various quantities on $\varepsilon$.

The basic requirement is that the ``would be" SUSY breaking
direction \( S \) as well as the orthogonal direction(s) \(T \)
are stabilized by the SNP superpotential \( W_{SNP} \) with
unbroken SUSY in the central region at some point
$(\widetilde{S}_0,\widetilde{T}_0)$. When the field theoretic
SUSY breaking is turned on, $(\widetilde{S}_0,\widetilde{T}_0)$
will be shifted by terms of $ O(\epsilon )$ to $({S}_0,{T}_0)$
\cite{foot1} and should satisfy the following criteria:\\  (a) a
minimum of the potential at \( ({S}_0,{T}_0) \): \( \pa
_{S}V|_{\rm min}=0 \),~\( \pa _{T}V|_{\rm min}=0 \) \\
(b) with broken SUSY: $ F_{S}|_{\rm min} \sim 0(\varepsilon)$ \\
 (c) and a small cosmological constant:
 $V|_{\rm min}<0(\varepsilon^2) $  \\
 (d) in the central region:
 \( ReS_{0},Re T_0\sim 1 \). \\
In addition, for the SUSY preserving direction $T$ we have\\
(e) \( F_{T}|_{\rm min}=0 \). \\
In the above \( |_{\rm min} \) means that the expression is to be
evaluated at the minimum of the potential. Similar conditions
apply for the complex conjugate quantities.

The potential is given in terms of the Kahler potential
$K=K(S,\bar{S},T,\bar{T})$, and the superpotential $W=W(S,T)$,
 \begin{equation}
 \label{potential}
 V=e^{-K}\left(
 F_{i}K^{i\bar{j}} F_{\bar j}-3|W|^{2} \right),
\end{equation}
 and the $F$ terms are given by
\begin{equation}
 \label{Fterm}
 F_{S}=\partial _{S}W+K_{S}W\, \, ;
 F_{T}= \partial_{T}W + K_{T}W.
\end{equation}
$K$ with subscripts denote derivatives of the Kahler potential,
and $K^{i\bar j}$ is the inverse of the matrix of second
derivatives of the Kahler potential. At the minimum the Kahler
metric is taken to be diagonal so that in particular
\begin{equation}
 \label{metric}
 K^{S\bar{T}}|_{min}=0.
\end{equation}

We will now show that in order to eventually obtain
parametrically small $O(\varepsilon )$ SUSY breaking the
superpotential and its first three derivatives must be of order
\( \varepsilon \) if the Kahler potential and its derivatives at
the minimum are of order unity, as expected. Conditions  (b),
(c) and (e) and the definition (\ref{Fterm}) imply that
\begin{equation}
\label{worders} W|_{\rm min},\ \partial _{S}W|_{\rm min},\
\pa_{T}W|_{\rm min}\sim O(\varepsilon ).
\end{equation}
Note also that  $ \pa _{\overline{S}}^{n}F_{S}|_{\rm min}=
\pa_{\overline{S}}^{n}K_{S}W|_{\rm min} \sim O(\varepsilon )$, and
 $ \pa_{\overline{S}}^{n}F_{T}|_{\rm min} =
\pa_{\overline{S}}^{n}K_{T}W|_{\rm min} \sim O(\varepsilon )$, for
$n=0,1,2, \dots$. Conditions (a),(b),(e) and eq.(\ref{metric})
then implies that
 $\partial _{S} F_{S}\partial _{S} K^{S\overline{S}} \bar{F}_{\bar S}+
O(\varepsilon ^{2})=0$, so that using (\ref{Fterm}) again we find
that $\partial _{S}F_{S}\sim O(\varepsilon )$ and hence $\partial
_{S}^{2}W|\sim O(\varepsilon )$. Also from (\ref{worders}) and the
second equation of  (a) we have \( \pa _{T}V|_{\rm
min}=e^{K}\pa_{T}F_{S}
 K^{S\overline{S}}F_{\overline{S}}+
 O(\varepsilon ^{2})=0\)
 leading to the estimate
 \begin{equation}
 \pa _{T}F_{S}|_{\rm min},\
 \partial_S\partial_TW|_{\rm min},\ \partial_SF_T|_{\rm min}\sim O(\varepsilon).
\end{equation}

Now, let us look for conditions on the second derivatives of the
potential. Using  conditions (a)-(e) and the above estimates we have,
\begin{eqnarray} \label{ss}
\partial ^{2}_{S}V|_{\rm min}= e^{K}
\Biggl[\partial^{2}_{S}F_{S}
K^{S\overline{S}}\overline{F}_{\overline{S}}|_{\rm min}+
O(\varepsilon^{2})\Biggr],
\end{eqnarray}
 so a priori $\partial ^{2}_{S}V|_{\rm min}\sim O(\varepsilon )$.
On the other hand using our previous estimates  we find that $\pa
_{S\overline{S}}V|_{\rm min}\sim O(\varepsilon^{2})$. Thus, the
subdeterminant $H_S$ of the matrix of second \( S \) derivatives
is given at the minimum by
\begin{eqnarray}\label{H}
 H_S|_{\rm min}&=&4\left[|\pa _{S\overline{S}}V|_{\rm min}|^{2}-
 |\pa_{SS}V|_{\rm min}|^{2}\right] \nonumber \\
 &=& 4\left[O(\varepsilon ^{4})-
 \left|e^{K}\pa_{S}^{2}F_{S}
 K^{S\overline{S}}\overline{F}_{\overline{S}}|_{\rm
 min}\right|^{2}\right].
 \end{eqnarray}
Now  $H_S|_{\rm min}$
must be positive. A necessary condition for this is
that  $\left|e^{K}\pa_{S}^{2}F_{S}
 K^{S\overline{S}}\overline{F}_{\overline{S}}|_{\rm
 min}\right|^{2}\sim O(\varepsilon^{2})$, which implies that
\(\pa_{S}^{2}F_{S}\sim O(\varepsilon ) \), and therefore that
$\pa_{S}^{3} W|_{\rm min}\sim O(\varepsilon )$.

Second $T$
derivatives are given by
\begin{eqnarray}
\pa _{T\overline{T}}V|_{\rm min} &=& e^{K} \pa_{T}F_{T}
K^{T\overline{T}} \pa_{\overline{T}} \overline{F}_{\overline{T}}
+ O(\varepsilon ^{}) \nonumber \\
\pa _{T}^{2}V|_{\rm min} &=& e^{K} 2\pa _{T}F_{T}K^{T\overline{T}}
\pa_{T}\overline{F}_{\overline{T}}+O(\varepsilon ).
\end{eqnarray}
But since \( \pa_{T} \overline{F}_{\overline{T}}
=\pa_{T}K_{T}\overline{W} \sim O(\varepsilon ) \), it follows
that $\pa _{T}^{2}V|_{\rm min}$  is  \( O(\varepsilon ) \).
Therefore the subdeterminant $H_T$ of \( T \) second derivatives
$H_T=4\left[|\pa _{T\overline{T}} V|_{\rm min}|^{2}-
|\pa_{TT}V|_{\rm min}|^{2}\right] $ can be positive with \(
\pa_{T}F_{T}=O(1) \).  This implies that (generically) $\pa_T^2 W\sim O(1)$,
and that $\pa^2_{T\bar{T}} V|_{min} \sim O(1)$.

Summarizing our results so far we find that the superpotential at
the minimum must satisfy
\begin{eqnarray}
\label{summary}
&& W|_{\rm min},~ \pa _{S}W|_{\rm min},
~\pa _{T}W|_{\rm min},~ \pa_{S}^{2}W|_{\rm min},
\nonumber \\ &&
\pa^2_{ST}W|_{\rm min},~ \pa _{S}^{3}W|_{\rm min}\sim
O(\varepsilon ).
\end{eqnarray}
On the other hand $\pa^2_T W$ and $\pa^4_S W$ and higher
derivatives in both $S$ (of order greater than 3) and in $T$
(greater than 1), as well as mixed derivatives of order greater
than 2 are generically of order unity. In the supersymmetric
limit $\varepsilon\rightarrow 0$, the landscape of the central
region looks very different in the $S$ and $T$ directions. In the
$T$ direction the potential $V$  is very steep, all derivatives
from the second derivative and up are generically of order 1 at
the minimum. In the $S$ direction the potential is flat around
the minimum. In particular the masses of the SUSY breaking $S$
moduli will be of order $\varepsilon$ while in general the masses
of the $T$ moduli will be of order one.


We now focus on the SUSY breaking direction $S$. First, at a high
scale all moduli are stabilized by SNP, and then at a lower scale
SUSY is broken. Our general considerations imply that $W_{SNP}$
has a fourth order zero at the minimum $\widetilde{S}_0$. Thus, the
simplest model for the SNP superpotential in the SUSY breaking
direction is $W_{SNP}=a_{4}(S-\widetilde{S}_{0})^{4}$, with $a_4,
\widetilde{S}_0 \sim O(1)$. The existence of an adjustable
constant is reflected in that the combination
$a_{4}\widetilde{S}_{0}^{4}$ is continuously adjustable even
though both $\widetilde{S}_{0}$ and $a_4$ are of order unity. To
realize the separate scale of stabilization of \( S \) and of
SUSY breaking there should be an additional small field theoretic
contribution \( \D W_{FT} \) (originating, for example, from
gaugino condensation) to the superpotential . The new SUSY
breaking minimum $S_0$ of the modified potential is shifted by a
small amount from the original minimum $\widetilde{S}_{0}$, and
at $S_0$ the cosmological constant vanishes (more precisely, it
is smaller than $\varepsilon^2$). We expect that all
non-perturbative field theoretic effects are proportional to
$e^{-1/g_{YM}^2}$, $g_{YM}$ being the unified coupling of the
field theory. $\cN=1$ SUGRA prescribes a chiral superfield whose
expectation value is $\langle S\rangle =1/g_{YM}^2$. It follows
that the SUSY breaking direction $S$ is in fact well approximated
by the very same direction. Note that $S$ could be very different
from any perturbative dilaton. In the central region it will be
some complicated function of the ten dimensional perturbative
dilaton and all the other perturbative moduli of any one theory
at a corner of the M-theory moduli space. Also we have not
specified what the Kahler potential of the field $S$ should be
beyond making the reasonable assumption that it and all its
derivatives are of order one.

In summary,  we expect  additional terms in the superpotential of
the form, $\Delta W_{FT}=\sum b_i e^{-\beta_i S}$. Now, this
correction term needs to be \( \Delta W \sim O(\varepsilon ) \),
to generate the hierarchy $m_{3/2}/M_P\sim 10^{-16}\sim e^{-40}$.
This implies that parameters $\beta_i$ (which are constants that
are determined, for example,  by the gauge group of some hidden
sector) need  to be \( \beta_i \sim 40 \) for prefactors \(
b_i\sim O(1) \). Alternatively, we may expand $\Delta W_{FT}$
around the  SUSY breaking  minimum, $ \Delta W_{FT}=\sum c_i
e^{-\beta_i (S-S_{0})}$, so $c_i \sim O(\varepsilon )$. In the
presence of $\Delta W_{FT}$ the modified superpotential in the
central region is of the following form
\begin{eqnarray}
\label{modelw}
  W=a_{4}(S-S_{0})^{4}+a_{3}(S-S_0)^{3}+
\nonumber \\
  a_{2}(S-S_{0})^{2}+a_{1}(S-S_{0})+a_{0}
\end{eqnarray}
In this expression all coefficients except for $a_4$ are of \(
O(\varepsilon ) \). The condition for a vanishing cosmological
constant becomes,
\begin{equation}
\label{finetuning}
 a_{1}=(\pm \sqrt{3K_{S\overline{S}}}-K_{S})a_{0}.
\end{equation}
This is a fine-tuning condition that almost certainly cannot be
satisfied if \( a_{0},a_{1} \) originate from the field theoretic
term $\Delta W_{FT}$ alone where the relation involves the
parameters $b_i$ and $\beta_i$ which are fixed by the field
theory.  Therefore, we argue that that \( a_{0} \) is a constant
coming from string theory that is able to adjust itself so as to
satisfy (\ref{finetuning}). Thus, in our model we can take,
$a_{4}\sim O(1)$, $a_{3}= - \frac{1}{6} \sum c_i \beta_i^3$,
$a_{2}= \frac{1}{2} \sum c_i \beta_i^2$, $a_{1}=-\sum c_i
\beta_i$ and \( a_{0} \) is an adjustable constant which
guarantees (\ref{finetuning}).

Let us  check that there is no obvious obstruction in this model
for having a SUSY breaking minimum by looking at the determinant
of second S-derivatives and showing that it can be positive. The
first term of \( H_S \)  is \( |\pa _{S\bar{S}}V|_{\rm min}|\sim
4|a_{2}|^{2}K^{S\overline{S}}|_{\rm min}=(\sum \beta_i^{2}c_i)^2
K^{S\overline{S}}|_{\rm min} \)  while in the second term $|\pa
_{S\bar{S}}V|_{\rm min}|$ the largest component \( \sim
6a_{3}a_{1}K^{S\overline{S}}|_{\rm min} = (\sum \beta_i^{3} c_i)
(\sum \beta_i c_i) K^{S\overline{S}}|_{\rm min} \). Thus the
positivity of \( H \) depends on the details of the model. We
have analyzed numerically various potentials, and found that
there is a range of parameters for which a minimum at $S_0$
exists. For example, a minimum at $S_0=1$ occurs if $a_1= 9 a_3$,
and $a_2= -\frac{3-\sqrt{3}}{4} a_1$.

The fact that $W_{SNP}$  has a fourth order zero (in the SUSY
limit) and therefore that all the coefficients up to and including
$a_3$ are $O(\epsilon)$ has an unwanted, but generic,
consequence: additional  minima which have a $O(\epsilon^2 )$
negative cosmological constant. Here we show how this comes
about. If both $F_S$ and $F_T$ vanish at some point then the
potential is negative, since at this point $V=-3|W|^2$. Of
course, there is no guarantee that there is a solution to these
conditions, however, we can argue reliably that near our SUSY
breaking minimum the potential becomes negative. This implies that
the minimum with vanishing cosmological constant is not the
global minimum. First, note that since we are interested in the
neighborhood of the point $(S_0,T_0)$ it is enough to keep the
terms in $W$ given in (\ref{modelw}). Let us fix $T=T_0$. Now
$F_S= \partial_S W+ K_S W\simeq 4 a_4(S-S_0)^3+ O(\varepsilon)$
has zeros at a distance of $(S-S_0)\sim \varepsilon^{1/3}$. At
this distance $W=W(S_0)+\pa_S W_{\rm min} (S-S_0)+\cdots$ is of
order $\varepsilon$, and
$F_T=(S-S_0)\partial_SF_T|_{min}+\cdots\sim\epsilon^{4/3}$ (see
(\ref{summary})) is of higher order. This means that at such
distances the potential is already negative. Thus, the SUSY
breaking minimum that we have identified cannot be a global
minimum. It is only a local minimum. A full discussion of the
 properties of these additional minima requires more
detailed information on the Kahler potential and the
superpotential and will not be attempted here. The appearance of
regions of negative potential is generic in $\cN=1$ SUGRA
theories, and their significance and implications should be
discussed in a cosmological context. We need to determine whether
the fields can move classically to the dangerous negative
potential regions, and whether they are stable if they are
already in the zero cosmological constant minimum. We will
discuss this issue separately.

Once we have obtained the SUSY breaking potential, which, for the
purpose of computing soft SUSY breaking terms,  could be well
approximated by a Polonyi type model with a linear superpotential
\( W=X+Y(S-{S}_{0}) \), we may follow the discussion of
\cite{KL,BIM} to obtain soft masses and relations between them.
However, to obtain actual spectra we need the coupling of matter
fields and gauge fields in the observable sector to moduli in the
central region. But these couplings are not available from
perturbative calculations and have to be constrained by some
general considerations. When (and if) expected results from the
Tevatron and LHC determine some of the SUSY soft terms, we may
use them to probe couplings in our effective theory.


In conclusion, the assumptions that we made were as follows:\\
i) The four dimensional low energy effective theory of strings is
a \( \cN=1 \) supergravity.\\
ii) All moduli are stabilized at around O(1) by stringy
non-perturbative effects which generate a superpotential for them
in the low energy theory. \\
iii) The superpotential and the Kahler potential coming from
stringy non-perturbative effects have only order 1 terms. \\
iv) Hierarchically small SUSY breaking effects are generated by
field theoretic effects at some intermediate scale.\\
v) The cosmological constant vanishes (or is small).\\
We have shown that\\
I) The superpotential generated by SNP effects must (in the
supersymmetric limit) have a fourth order zero in the
supersymmetry breaking direction $S$, and needs to have only a
second order zero
in the SUSY preserving directions $T$.\\
II) The potential  is very steep in the $T$ directions while it
is very flat in $S$ directions near the minimum. Consequently,
SUSY breaking moduli have electroweak scale masses, while in
general SUSY preserving moduli will have string scale masses. \\
III) In order to ensure a vanishing or small cosmological constant
there must be an adjustable constant in the superpotential
originating directly from string theory that tracks the low
energy SUSY breaking effects \cite{foot2}.

\acknowledgements

We thank I. Antoniadis, J. Lykken, B. Ovrut and S. Kachru for
useful discussions. This research was partly supported by grant
No.~1999071 from the United States-Israel Binational Science Foundation
(BSF), Jerusalem, Israel.  SdA was also supported in part by the
United States Department of Energy under grant
DE-FG02-91-ER-40672. SdA wishes to acknowledge the hospitality of
the ITP Santa Barbara during the final phase of this work.


\begin{thebibliography}{99}
\bibitem{foot0}
The resolution of the cosmological constant problem is extremely
important, but is outside the scope of our discussion.
\bibitem{foot1}
The smallness of the gauge coupling
(at unification $g^2_{YM}\sim 1/25$) is not incompatible with moduli stabilized
at O(1) as discussed in detail in \cite{bda}.
\bibitem{foot2}
Assuming that there is no
hitherto unknown field theoretic mechanism that does this.
\end{thebibliography}
\end{document}